\documentclass[fleqn,twoside]{article}
\usepackage{graphics,graphicx}
\usepackage{gc,amsfonts,amssymb,cite}

\def\mn{_{\mu\nu}}
\def\MN{^{\mu\nu}}

\def\cK{{\cal K}}
\def\cV{{\cal V}}

\def\kappa{\varkappa}

\def\wt{\widetilde}
\def\tg{{\wt g}}
\def\tR{{\wt R}}

\def\M{{\mathbb M}}

\def\Lambdaef{\Lambda_{\rm eff}}
\def\sss{\scriptscriptstyle}

\def\mD{m_{\sss D}}

\begin{document}
\twocolumn[
\Arthead{13}{2007}{4 (52)}{1}{6}

\Title{Abilities of multidimensional gravity}

\Authors{K.A. Bronnikov\foom 1} {and S.G. Rubin\foom 2}
    {Centre for Gravitation and Fundamental Metrology, VNIIMS,
    46 Ozyornaya St., Moscow, Russia;\\
    Institute of Gravitation and Cosmology, PFUR,
    6 Miklukho-Maklaya St., Moscow 117198, Russia}
    {Moscow Engineering Physics Institute, 31 Kashirskoe Shosse,
    Moscow 115409, Russia}


\Abstract
 {We show that a number of problems of modern cosmology may be addressed
  and solved in the framework of multidimensional gravity with high-order
  curvature invariants, without invoking other fields. As applications of
  this approach, we mention primordial inflation and particle production
  after it; description of the modern accelerated stage of the Universe with
  stable compact extra dimensions; construction of asymmetric thick
  brane-world models.}


] 

\email 2 {kb20@yandex.ru}
\email 3 {sergeirubin@list.ru}

\section{Introduction}


  We consider multidimensional gravity as a basis for solving many
  fundamental problems using a minimal set of postulates. The particular
  value of the total space-time dimension $D > 4$ and the topological
  properties of space-time are supposed to be determined by quantum
  fluctuations and may vary from one space-time region to another, leading
  to drastically different universes. It turns out that different effective
  theories take place even with fixed parameters of the original Lagrangian.

  As a result, the situation to a certain extent resembles the prediction
  of string theory known as the landscape concept: the total number of
  different vacua in heterotic string theory is about $10^{1500}$
  \cite{Landscape0,Landscape1}). The number of possible different universes
  is then huge but finite. Moreover, the concept of a random potential
  \cite{Random}, also referring to quantum fluctuations, leads to an
  infinite number of universes with various properties. We make a step
  further and try to ascribe the origin of such potentials to
  multidimensional curvature-nonlinear gravity. We then use the obtained
  effective theories, containing scalar fields of purely geometric origin,
  to a number of well-known problems of modern cosmology.


  Some challenging problems are those related to fine tuning, required to
  explain the actual properties of our Universe. Thus, small parameters are
  necessary to provide the smallness of temperature fluctuations of the CMB;
  an extremely small parameter is required to explain the observed value of
  dark energy density. It can be shown, in particular, that such fine-tuning
  problems may be reduced to the problem of the number of extra dimensions
  \cite{BR06}.

\section{$F(R)$ gravity, a single factor space}

\subsection{Basic equations}

    Let us first consider, for simplicity, $D$-dimensional space-time with
    the structure $\M_D = \M_d \times \M_4$, where the extra factor space
    $\M_d$ is of arbitrary dimension $d$ and is assumed to be a space of
    positive or negative constant curvature $k = \pm 1$. Consider the action
\beq                                                    \label{S1}
    S = \Half \mD^{D-2} \int \sqrt{^{D}g}\, d^{D}x\, [F(R) + L_m]
\eeq
    and the $D$-dimensional metric
\beq                                                            \label{ds_D}
     ds^{2} = g_{\mu\nu}(x) dx^{\mu}dx^{\nu}
                        + \e^{2\beta(x)}h_{ab}dx^{a}dx^{b}
\eeq
    where $(x)$ means the dependence on $x^\mu$, the coordinates of $\M_4$;
    $h_{ab}$ is the $x$-independent metric in $\M_d$. The choice
    (\ref{ds_D}) is used in many studies, e.g., \cite{Zhuk,Holman,Majumdar}.

    Capital Latin indices cover all $D$ coordinates, small Greek ones cover
    the coordinates of $\M_4$ and $a, b, \ldots$ the coordinates of $\M_d$.
    The $D$-dimensional Planck mass $\mD$ does not necessarily coincide with
    the conventional Planck scale $m_4$; $\mD$ is, to a certain extent, an
    arbitrary parameter, but on observational grounds it must not be smaller
    than a few TeV.

    The Ricci scalar can be written in the form
\bear                                                  \label{R-decomp}
        R \eql R_{4} + \phi + f_{\rm der},
\nn
        \phi \eql kd(d-1) \mD^2 \e^{-2\beta(x)}
\nn
      f_{\rm der} \eql 2d g^{\mu\nu}\nabla_{\mu}\nabla_{\nu}\beta
                + d(d+1) (\d\beta)^2,
\ear
    where $(\d\beta)^2 = g\MN \d_\mu \beta \d_\mu \beta$. The {\sl
    slow-change approximation\/}, suggested in \cite{BR06}, assumes that all
    quantities are slowly varying, i.e., it considers each derivative
    $\d_{\mu}$ (including those in the definition of $R_4$) as an expression
    containing a small parameter $\eps$, so that
\beq
        |\phi| \gg |R_4|,\ |f_{\rm der}|.
\eeq
    As shown in \cite{BR06}, this approximation even holds in any
    inflationary model whose characteristic energy scale is far below the
    Planck scale $\mD$, to say nothing of the modern epoch.
    Thus, the GUT scale, which is common in inflationary models, is $m_{\rm
    GUT} \sim 10^{-3} m_4$, which means that primordial inflation may
    be well described in the present framework if $\mD \sim m_4$.

    In this approximation, using a Taylor decomposition for $F(R) = F(\phi +
    R_{4} + f_{\rm der})$, integrating out the extra dimensions and using a
    conformal mapping to pass over to a 4-dimensional Einstein frame
    with the metric $\tg\mn$, we obtain up to $O(\eps^2)$:
\bear
     S \eql \frac{\cV[d]}{2} \mD^2 \int d^{4}x\, \sqrt{\tg}\, (\sign F') L,
\nn
     L \eql \tR_4 + \Half K_{\rm Ein}(\phi) (\d\phi)^2
                        - V_{\rm Ein}(\phi) + {\wt L}_m,   \label{Lgen}
\nnn \cm
          {\wt L}_m = (\sign F')\frac{\e^{-d\beta}}{F'(\phi)^2} L_m;
\\
     K_{\rm Ein}(\phi) \eql                                \label{KE}
        \frac{1}{2\phi^2} \biggl[
            6\phi^2 \biggl(\frac{F''}{F'}\biggr)^2\!
            -2 d \phi \frac{F''}{F'} + \Half d (d + 2)\biggr]\,,
\nnn
\\
     V_{\rm Ein}(\phi) \eql - (\sign F')
        \left[\frac{|\phi|\mD^{-2}}{d (d -1)}\right]^{d/2}
                \frac{F(\phi)}{F'(\phi)^2 }.                \label{VE}
\ear
    In (\ref{Lgen})--(\ref{VE}), the tilde marks quantities obtained from or
    with $\tg\mn$; the indices are raised and lowered with $\tg\mn$;
    everywhere $F = F(\phi)$ and $F' = dF/d\phi$. All quantities of orders
    higher than $O(\eps^2)$ are neglected.

    The quantity $\phi$ may be interpreted as a scalar field with the
    dimensionality $\mD^2$, see (\ref{R-decomp}). In what follows (if not
    indicated otherwise) we put $\mD =1$.

    In fact, the function $F(\phi)$ represents an infinite power series
    inevitably caused by quantum corrections. Here we will
    use the truncated form
\beq                            \label{quartic}
        F(R) = R + cR^{2} + w_{1}R^{3} + w_{2}R^{4} - 2\Lambda
\eeq
    and demonstrate that some new nontrivial results may be obtained even
    under such simplified assumptions. Curvature-quartic multidimensional
models were also studied in \cite{Zhuk8}.

\subsection{Some applications}

    For certainty, we will everywhere treat the Einstein conformal
    frame, in which the Lagrangian has the form (\ref{Lgen}), as the
    physical frame used to interpret the observations. This assumption is
    not inevitable and is even rather arbitrary; the choice of a
    conformal frame is known to be related to the properties of the set of
    measurement instruments used \cite{stan-mel,bm-erice}, and using other
    conformal frames is also possible but goes beyond the scope of this
    paper.

\subsection* {Effective particle production after inflation}

    According to \cite{DolgovAbbott}, quick oscillations of the inflaton
    field immediately after the end of the inflationary stage are necessary
    for particle production which should lead to the observable amount of
    matter. It is known \cite{KLS, Shtanov} that if the inflaton coupling
    to matter fields is negligible, the mechanism of particle/entropy
    production and heating of the Universe could be ineffective. On the
    other hand, a strong coupling between the inflaton and matter fields,
    leading to large quantum corrections to the initial Lagrangian, would
    set to doubt the sufficiently small values of the input parameters
    needed for the very existence of the inflationary stage.

    Another possibility of effective particle creation (parametric
    resonance) was described in \cite{KLS}. Nonlinear multidimensional
    gravity yields one more mechanism \cite{brost07}.

    Consider the potential and kinetic terms for the effective Lagrangian
    (\ref{Lgen}) with the function $F(R)$ taken from (\ref{quartic}).
    A complicated $\phi $ dependence of the kinetic term strongly affects
    the classical scalar field dynamics. The effect is especially strong
    when the minimum of the kinetic term approximately coincides with that
    of the potential. To illustrate the situation, consider a toy model of
    a scalar field $\phi$ with
\bear                                       \label{toy}
    V(\phi ) \eql \frac12 m^2\phi^2
\nn
         K(\phi) \eql K_1 (\phi - \phi_{\min})^2 + K_{\min},
           \quad  K_1,\ K_{\min} >0.
\ear
    The field equations then read:
\bearr                                                        \label{fr-eq}
    H^{2} =\frac{\kappa^{2}}{3}\left[
        \frac{1}{2}K(\phi)\dot{\phi}^{2} + V(\phi)\right]  ,
\nnn
    K(\phi)\left(\ddot{\phi} + 3H\dot{\phi}\right)
        + \frac{1}{2}K_{\phi}(\phi)\dot{\phi}^{2}
            + V_{\phi} (\phi) = 0
\ear
   (the index $\phi$ means $d/d\phi$, and $H$ is the Hubble parameter).

   When the inflation is over, the amplitude of inflaton oscillations is
   small at the Planck scale, and the effective kinetic term $K_{\rm
   eff}\sim K_{\min}$ is small due to the chosen values of the parameters.
   The effective Lagrangian, containing an interaction term of the inflaton
   and some other scalar field $\chi$
\beq                                \label{Leff}
    L_{\rm eff}\simeq \frac12 K_{\min}\dot{\phi}^2 + g\phi\chi\chi,
\eeq
   can be brought into the standard form by re-definition of the
   inflaton field:
\beq                                \label{LeffStand}
    L_{\rm eff}\simeq\frac12 \dot{\phi}^2 +
            \frac{g}{\sqrt{K_{\min}}}\phi\chi\chi.
\eeq
   A small value of $K_{\min}$ increases the effective coupling constant
   (by an order of magnitude for the chosen values of the parameters) and
   surely leads to more rapid particle production. As a result, we overcome
   the above difficulty: a slow motion during inflation may be reconciled
   with effective particle production right after inflation. Hence
   universes of this sort possess promising conditions for creation of
   complex structures.

\subsection* {Effect of the number and geometry of extra dimensions}

   Even if the initial Lagrangian is entirely specified, low-energy
   effective Lagrangians, drastically different from one another, can be
   obtained by varying $d$ and the curvature index $k$ of the extra
   dimensions.

   A strong influence of the number $d$ on the effective Lagrangian
   parameters is evident since the potential (\ref{VE}) contains a
   quickly decreasing factor $\sim d^{-d}$. Thus, if in a stationary state
   $\phi = \phi_0$, the dimensionless initial parameters $|\phi_0|\mD^{-2}$
   and $F'(\phi_0)$ are of order unity, the effective cosmological constant
   $\Lambdaef = V_{\rm Ein}(\phi_0)$ is related to $F(\phi_0)$ (which may
   be close to $\mD^2$) by
\beq
    \Lambdaef/F(\phi_0) \sim [d(d-1)]^{-d/2}.
\eeq
   It is of interest that $d^{-d} \approx 10^{-123}$ for $d = 67$. Thus, at
   least in the Einstein picture, a fluctuation leading to a
   (67+4)-dimensional space may evolve to a space with the vacuum energy
   density $10^{-123}\,m_4$. The extreme smallness of $\Lambdaef$ is
   related to the number of extra dimensions, and other physical ideas are
   not required. We have taken for certainty $\mD = m_4$, otherwise
   the estimates will be slightly different.

   Fig.\,1 gives another example of $d$-dependence of the shape of the
   potential.  Even its minimum does or does not exist depending on the
   value of $d$, see the curves at $\phi > 0$.  If a universe is nucleated
   with extra dimensions having negative curvature, we have $\phi < 0$, see
   Fig.\,1.  Evidently the mean value of the potential $V(\phi )$ in such a
   universe tends to infinity for $d=2$, to a constant value for $d=4$ and
   to zero for $d=6$.  All this takes place if the initial field value is
   less than $-1$. Otherwise, if a universe is born with $-1<\phi < 0$, it
   is captured in a local minimum, and the size of extra dimensions remains
   small.

\EFigure{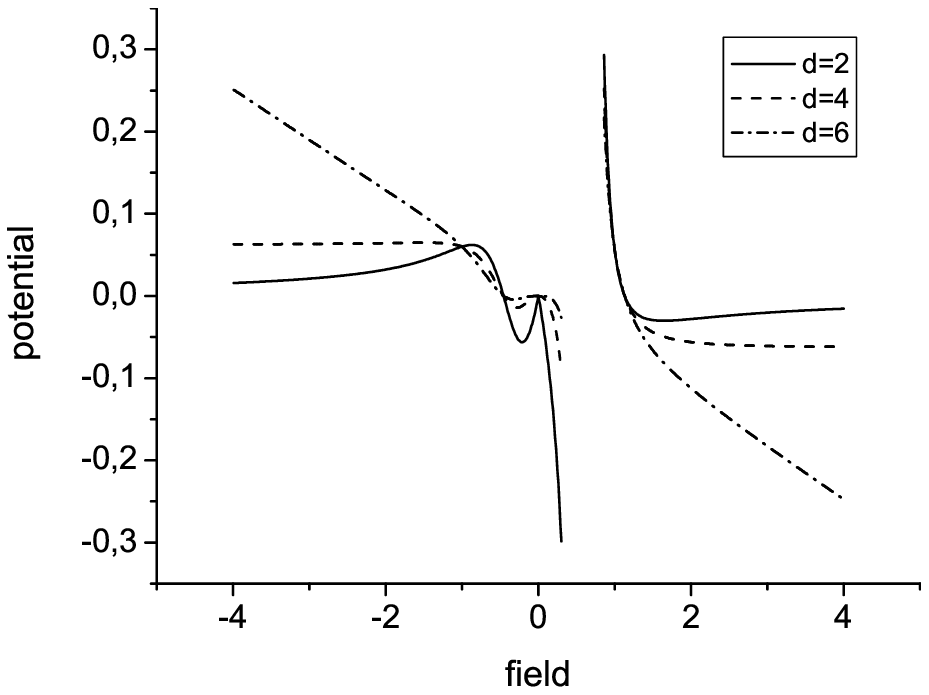}    
    {Different potentials for different $d$, obtained from the
    Lagrangian (\ref{Lgen}) with $F(R)$ in the form (\ref{quartic}).
    Parameters: $c=0,\ w_1 =0,\ w_2 =-1,\ \Lambda =-0.25$.
    The curves are adapted to a unique scale.} \label{d=2-6a}

\section {Multiple factor spaces}

   So far we discussed effective low-energy theories corresponding
   to the metric (\ref{ds_D}) for different choices of the initial action.
   Some values of the parameters turned out to be suitable for the
   description of a universe like ours. Additional opportunities appear when
   varying the structure of extra dimensions, which includes the number of
   extra factor spaces, their dimensionality and curvature.

   As was mentioned in the Introduction, we do not assume a specific number
   of extra dimensions or their topology. Both are thought to arise due to
   quantum fluctuations close to the Planck scale. If quantum fluctuations
   lead to a more complex structure of the extra dimensions, the physics
   becomes much richer.

\subsection {Two compact extra factor spaces}

   Consider an extra space being a product of two factor spaces: $\M_{d} =
   \M_{d_1} \times \M_{d_2}$, with the metric

\beq                                                            \label{ds_D2}
     ds^2 = g_{\mu\nu}(x) dx^{\mu}dx^{\nu}
                        + \sum_i \e^{2\beta_i(x)} h^{(i)}_{ab}dx^{a}dx^{b},
\eeq
   where the index $i = 1,2$ enumerates the extra factor spaces and
   $h^{(i)}_{ab}$ have the same properties as $h_{ab}$ in the previous
   section, with the curvature indices $k_i = \pm 1$ similar to $k$ in
   (\ref{R-decomp}). Considering the same action (\ref{S1}), we should
   introduce two scalar fields
\beq
        \phi_i = k_i (d_i-1) \mD^2 \e^{-2\beta_i(x)}
\eeq
   to describe the low-energy limit. The effective potential has the
   following form in the Einstein frame:
\bearr
    V_{\rm Ein}(\phi_1 ,\phi_2 )
\nnn
    = -\frac12 \frac{\big(\sign F'(\phi)\big)}{[F'(\phi)]^2}
        \frac{|\phi_1|^{d_1/2}}{ [(d_1 {-} 1)]^{d_1 /2}}
            \frac{|\phi_2|^{d_2/2}}{ [(d_2 {-} 1)]^{d_2/2}}
\nnn  \nhq
    \times \biggl[ F(\phi)
    + d_1 \phi_1^2 \biggl(\! c_1 + \frac{2c_2}{d_1{-}1} \!\biggr)
    + d_2 \phi_2^2 \biggl(\! c_1 + \frac{2c_2}{d_2{-}1}\!\biggr)\biggr],
\nnn                            \label{VE2fields}
\ear
   where $\phi = d_1 \phi_1 + d_2 \phi_2$.
%
\EFigure{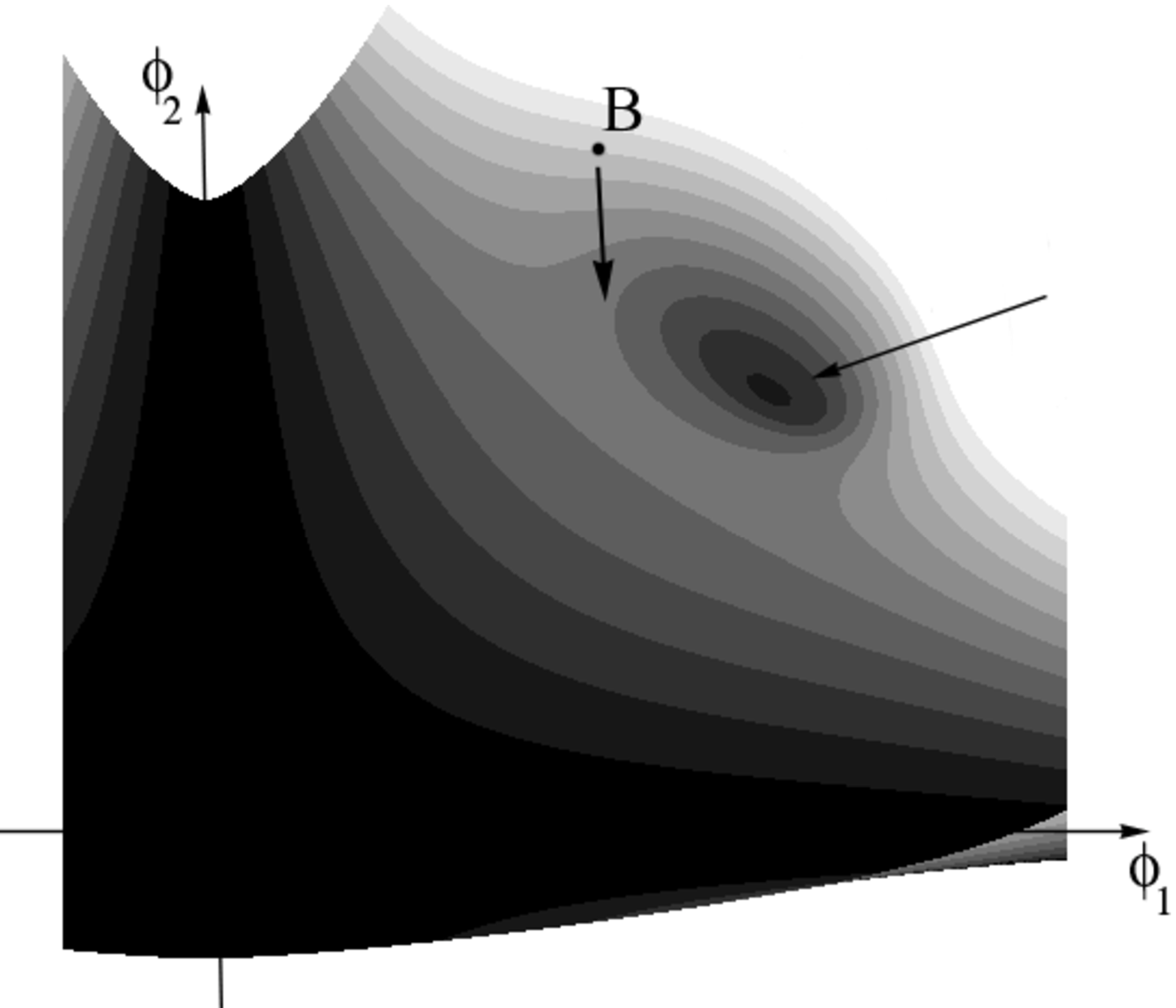}  
    {Effective potential for an extra space with the topology
    $\M_d = \M_{d_{1}}\times \M_{d_{2}}$, $d_1 = d_2 = 3$, and the parameters
    are $c=-0.5,\ \Lambda =0.2,\ c_1 = c_2 =-0.38$.  The view is from above,
    lower levels are darker. The local minimum is marked by a long arrow.}
    \label{TwoFields}
%
   It is presented in Fig.\,\ref{TwoFields} where two valleys of the
   potential lie in perpendicular directions, $\phi_1 =0$ and $\phi_2 =0 $,
   each of them corresponding to an infinite size of one of the extra
   factor spaces, $\M_{d_1}$ or $\M_{d_2}$. Of greater interest is the local
   minimum, marked by a long arrow, where both factor spaces are compact and
   have a finite size. A universe can live long enough in this metastable
   state, as in the case of a simpler topology of extra dimensions
   discussed above.

   An interesting possibility arises if a universe is formed at point B in
   Fig.\,2. There occurs inflation, and it ends as the fields move from
   point B along the arrow. The fate of different spatial domains depends on
   the field values in these domains. Even if most of them evolve to the
   metastable minimum, some part of the domains overcome the saddle and
   tend to the first or second valley, with an infinite size of one of the
   factor spaces. In this case, our Universe should contain some domains of
   space with macroscopically large extra dimension. Their number and size
   crucially depend on the initial conditions.

   The laws of physics in such a domain are quite different from ours. So,
   if, say, a star enters into such a domain, since the law of gravity is
   dimension-dependent, the balance of forces inside the star will be
   violated, and it will collapse or decay. More than that, there will be no
   usual balance between nuclear and electromagnetic forces in the stellar
   matter, so that even nuclei (except maybe protons) will decay as well.
   Even hadrons, being composite particles, are likely to lose their
   stability.

\subsection {Thick asymmetric brane world models}

   Quite a different configuration is obtained if, again using the metric
   (\ref{ds_D2}), we assume that one of the extra factor spaces is
   noncompact, thus considering a mixed ansatz: let, say, $\M_{d_1}$ be
   noncompact and, for simplicity, one-dimensional ($d_1=1$) while
   $\M_{d_2}$ is, as before, a compact factor space in the spirit of the
   Kaluza-Klein concept. As a result of the reduction procedure described
   above, we obtain (in the Einstein frame) an effective 5D theory
   in the space-time $\M_5 = \M_4 \times \M_{d_1=1}$ with the Lagrangian
   \cite{BR07}
\beq                                                          \label{L5}
      L_{\rm eff} = R_5 + K(\phi) g^{ab}\d_a\phi \d_b\phi -2V(\phi),
\eeq
   where $R_5$ is the 5D Ricci scalar, $\phi$ is an effective scalar field,
   obtained in the above manner from the scale factor of $\M_{d_2}$, and we
   have chosen the units so that the effective 5D gravitational constant is
   equal to unity. The indices $a,b$ cover all five coordinates. The
   functions $K(\phi)$ and $V(\phi)$ are obtained from the initial
   $D$-dimensional action similarly to the quantities $K_{\rm Ein}$ and
   $V_{\rm Ein}$ in \eqs (\ref{KE}), (\ref{VE}). The 5D metric is chosen in
   the form
\beq                                                       \label{ds5}
     ds^2 = \e^{2\gamma (y)}dt^2
              - \e^{2\beta (y)} d{\vec x}{}^2 - dy^2,
\eeq
   where $\{\vec x, t\}$ are the usual 4D coordinates and $y$ is the fifth
   coordinate. Assuming global regularity, we seek models of thick brane
   worlds able to trap ordinary matter in some layer described by a small
   range of the coordinate $y$.

   Many models of interest can be obtained if we use a more general initial
   action than (\ref{S1}), namely,
\bearr \nhq                                                 \label{S_D}
     S =  \Half \mD^{D-2}\!\int\! \sqrt{^D g}\,d^D x \Big[ F(R)
\nnn \inch\cm
        + c_1 R_{AB} R^{AB} + c_2 \cK \Big],
\ear
   where $R_{AB}$ is the Ricci tensor and $\cK = R_{ABCD}R^{ABCD}$ is the
   Kretschmann scalar in $\M_D$; $c_1$ and $c_2$ are constants, and we take,
   as before, the function $F(R)$ in the form (\ref{quartic}). It should be
   noted that the reduction procedure, involving the slow-change
   approximation, works well with this more general action (and even more
   complex ones) and leads to effective 5D Lagrangians of the form
   (\ref{L5}).

\EFigure {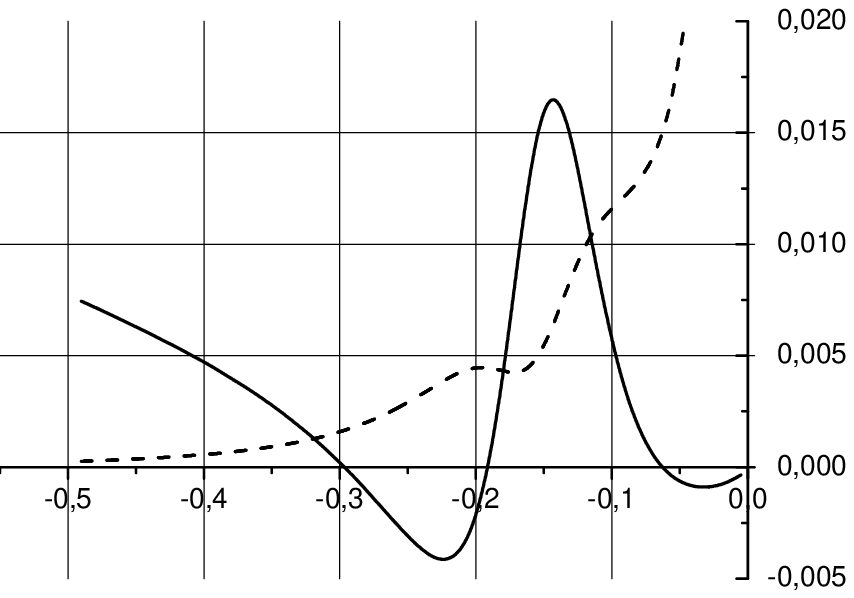}
   {The parameter values given in \eq(\ref{param2}). The solid
    line shows the potential, the dashed line represents the kinetic
    term multiplied by the factor $10^{-5}$.}\label{V_RS2}
   Fig.\,2 presents an example of the functions $V(\phi)$ and $K(\phi)$
   corresponding to the parameters
\bearr                                                \label{param2}
    d_0= 5,\ \ d_1 =2,\ \ c=6,\ \ w_1=15,\ \ w_2 = 7,
\nnn \cm
        \Lambda = -0.02,\ \ c_k = 4(c_1+c_2)=5,
\nnn \cm
    c_v = c_1 + 2c_2/(d_2-1) =1.5,
\ear
   which give rise to an asymmetric thick brane in 5D space, similar in its
   main features to the second Randall--Sundrum model (RS2) \cite{RS2}.
   The corresponding solution to the field equations that follow from
   (\ref{L5}) describes a space-time with the metric (\ref{ds5}) which is
   asymptotically anti-de Sitter (AdS) at large $|y|$, with the decressing
   warp factor $\e^{\gamma} = \e^{\beta} \sim \e^{-k_{\pm}|y|}$, though with
   different curvature values as $y\to +\infty$ and $y\to -\infty$
   ($k_+\ne k_-)$; the scalar field $\phi$ has the conventional form of a
   kink interpolating between two minima of the potential while $K(\phi)$
   remains positive. Like the RS2 branes, these configurations have AdS
   horizons at both sides far from the brane; they are able to trap a
   massless scalar field and, since gravitational perturbations are known to
   behave as a massless scalar \cite{wolfe}, such models naturally provide
   the validity of Newton's law with proper corrections \cite{RS2,sas03} for
   matter on the brane. Assuming that the brane asymmetry is negligible,
   approximate expressions for the modified Newtonian potential $V_N$ for
   gravity on the brane at large and small radii $R$ are \cite{call04}
\bearr
     V_N (R) \approx -\frac{GM}{R} \biggl( 1 + \frac{2l_c}{3R}  \biggr),
            \cm\ R \ll l_c,                       \label{N-approx}
\nnn
     V_N (R) \approx -\frac{GM}{R} \biggl(1 + \frac{2l_c}{3R^2}\biggr),
            \cm R \gg l_c.
\ear
   where $l_c$ is the mean AdS curvature radius. At radii $\gtrsim 10\,l_c$,
   the potential almost coincides with the Newtonian one while at small
   radii as compared with $l_c$ (but large as compared with the brane
   width) gravity becomes effectively 5-dimensional. The Newtonian
   gravitational constant $G$ is related to the 5D Planck mass $m_5$ as
   follows:
\beq
          8\pi G = (m_5^3 l_c)^{-1},                        \label{G-RS2}
\eeq
   where $m_5$ is connected with the initial multidimensional Planck
   mass $\mD$ by a coefficient of order unity.

   A well-known problem with such branes is their inability to trap massive
   scalar fields \cite{br-meier}.

   A peculiar feature of another class of BW models obtained from (\ref{L5})
   (with other sets intial parameters) is that $K(phi)$ has a variable sign,
   i.e., the scalar kinetic energy is positive in some range of $\phi$
   values and negative in another range. The brane is located in the
   vicinity of the transition point between these ranges and is thus highly
   asymmetric by construction. In such BW models, the warp factor again
   corresponds to AdS space far from the brane, but exponentially grows,
   which provides a purely gravitational mechanism for matter field
   trapping, without need for any interaction between $\phi$ and matter
   fields \cite{br-meier}.  There is a problem with massless field and
   graviton trapping since it turns out that the discrete spectrum of
   trapped scalar field modes does not contain a massless mode.  These
   branes are, however, similar to one of the two branes in the RS1 model
   \cite{RS1}, namely, the one with negative tension, which is interpreted
   as the observable (Planck) brane, and Newton's law with certain
   corrections is known to hold there \cite{RS1,smol02,arn04}. Our models
   can be interpreted in terms of an RS1-like structure, with a circular
   extra dimension, but with a positive-tension brane moved to infinity. In
   this case, a viable law of gravity on the brane can be obtained and may
   be written as \cite{BR07}
\bearr
     V_N (R) \approx -\frac{G_1 l_c M}{3\pi R^2}, \cm R \ll l_c,
\nnn                                                    \label{N-approx1}
     V_N (R) \approx -\frac{G_1 M}{3 R},    \cm\ R \gg l_c.
\ear
   where $8\pi G_1 = 1/(m_5^3 l_c)$. Thus, for $R \gg l_c$, Newton's law is
   valid with the gravitational constant $G = G_1/3$ (whereas in
   (\ref{N-approx}) we had $G = G_1$); at small radii, as before, the
   potential is proportional to $R^{-2}$, though with another coefficient.

   The AdS curvature radius $l_c$ remains an arbitrary parameter in all such
   models.

\section{Conclusion}
\label{Conclusion}

   We have discussed the ability of nonlinear multidimensional gravity
   to produce various low-energy effects and models, some of which could
   describe our Universe. We only assume a certain form of the pure
   gravitational action and a sufficient number of extra dimensions but do
   not fix the number, dimensions and curvature signs of extra factor
   spaces. Artificial inclusion of matter fields is not supposed, so that
   our conclusions are based on purely geometric grounds.

   We have seen that the same theory described by a specific Lagrangian
   leads to a diversity of low-energy situations, depending on the structure
   of extra dimensions and the initial conditions.

   We have found out, in particular, that nontrivial forms of the kinetic
   term of the inflaton field arise naturally in this approach. As a result,
   inflaton oscillations at the end of inflation could be very rapid with an
   appropriate form of the kinetic term. This increases the particle
   production rate after the end of inflation. At the same time, a slow
   motion of the inflaton at the beginning of inflation provides a
   sufficiently long inflationary stage.

   It has also been shown that multiple production of closed walls and hence
   massive primordial black holes is a probable consequence of modern models
   of inflation \cite{PBH}.

   Quite different effective low-energy models arise if one considers
   different numbers and/or topology of extra dimensions, even if all
   parameters of the initial Lagrangian are fixed. It means that the
   specific values of these parameters could be less important than it is
   usually supposed: even more important are the number, dimensions and
   curvatures of the extra factor spaces. In particular, varying the
   number $d$ of extra dimensions forming a single factor space, it is quite
   easy to obtain the proper value of the inflaton mass.

   We have seen that the size of extra dimensions may depend on the spatial
   point in the observed space, so that our Universe may contain spatial
   domains with a macroscopic size of extra dimensions, where the whole
   physics should become effectively multidimensional.

   We have also obtained 5D gravitational kinky configurations which could
   be interpreted as thick branes in the spirit of the widely discussed
   brane world concept.

   Thus pure multidimensional gravity, even without other ingredients,
   is quite a rich structure, and many problems of modern cosmology may be
   addressed in this framework. A task of interest is to try to construct
   a model able to solve a number of such problems (if not all)
   simultaneously.

\Acknow
{K.B. acknowledges partial financial support from Russian Basic Research
Foundation Projects 05-02-17478 and 07-02-13614-ofi-ts.}

\end{document}